\documentclass[12pt,preprint]{aastex}

\newcommand{\be}{\begin{equation}}
\newcommand{\e}{\end{equation}}
\newcommand{\f}{\frac}

\shorttitle{KSZ anisotropies generated by gas in cosmic structures}

\begin{document}
\title{Kinematic Sunyaev-Zeldovich Cosmic Microwave Background Temperature
Anisotropies Generated by Gas in Cosmic Structures.}

\author{F. Atrio-Barandela\altaffilmark{1},
J.P. M\"ucket\altaffilmark{2}, R. G\'enova-Santos\altaffilmark{3}}
\altaffiltext{1}{F\'{\i}sica Te\'orica, Universidad de Salamanca, 37008 Salamanca, Spain;
email: atrio@usal.es}
\altaffiltext{2}{Astrophysikalisches Institut Potsdam.  D-14482 Potsdam, Germany;
email: jpmuecket@aip.de}
\altaffiltext{3}{Cavendish Laboratory.  University of Cambridge. Cambridge CB3OHE, U.K;
email: rgs@mrao.cam.ac.uk}

\begin{abstract}
If the gas in filaments and halos shares the same velocity field
than the luminous matter, it will generate measurable temperature anisotropies
due to the Kinematic Sunyaev-Zeldovich effect. We compute the distribution
function of the KSZ signal produced by a typical filament and show it is highly
non-gaussian.  The combined contribution of the Thermal and
Kinematic SZ effects of a filament of size $L\simeq 5$Mpc and electron
density $n_e\simeq 10^3m^{-3}$ could explain the cold spots of $\delta\sim -200\mu$K
on scales of $30'$ found in the Corona Borealis Supercluster by the VSA experiment.
PLANCK, with its large resolution and frequency coverage,
could provide the first evidence of the existence of filaments
in this region. The KSZ contribution of the network of filaments and halo
structures to the radiation
power spectrum peaks around $l\sim 400$, a scale very different from that of 
clusters of galaxies, with a maximum amplitude 
$l(l+1)C_l/2\pi\sim 10-25 (\mu K)^2$, depending on model parameters, i.e.,
$\sigma_8$ and the Jeans length. About 80\% of the signal comes from filaments
with redshift $z\le 0.1$. Adding this component to the intrinsic
Cosmic Microwave Background temperature anisotropies of the concordance model
improves the fit to WMAP 3yr data by $\Delta\chi^2\simeq 1$. The improvement is 
not statistically significant but a more systematic study could demonstrate
that gas could significantly contribute to the anisotropies measured by WMAP.
\end{abstract}


\section{Introduction}
At low redshift, about 50 \% of all baryons have not been identified
(Fukugita \& Peebles, 2004). The highly ionized intergalactic gas, that has evolved 
from the initial density perturbations into a complex
network of mildly non-linear structures in the redshift interval $0<z<6$
(Rauch 1998; Stocke, Shull \& Penton 2004),  could contain most of the 
baryons in the Universe (Rauch et al. 1997, Schaye 2001, Richter et al 2005). 
With cosmic evolution, the fraction 
of baryons in these structures decreases as more matter is concentrated within 
compact virialized objects. The Ly$\alpha$ 
forest absorbers at low redshifts are filaments (with low HI column densities) 
containing about 30\%  of all baryons
(Stocke et al. 2004). Hydrodynamical simulations predict that 
another large fraction of all baryons resides within mildly-nonlinear 
structures with temperatures $0.01-1$KeV, called  Warm-Hot Intergalactic
Medium (WHIM). The amount of baryons within this WHIM is estimated to reach
20 - 40\% (Cen \& Ostriker 1999; Dave et al. 1999, 2001). 

Being highly ionized the WHIM could generate temperature anisotropies 
by means of the thermal or kinematic Sunyaev-Zeldovich effect 
(TSZ, KSZ; Sunyaev \& Zeldovich 1972, Sunyaev \& Zeldovich 1980),
adding to the contribution coming from gas in halos. 
Hallman et al (2007) carried out a numerical simulation
and found that about a third of the SZ flux would be generated by unbound gas.
In Atrio-Barandela \& M\"ucket (2006) -hereafter ABM- we estimated the CMB anisotropies
generated by the WHIM due to the TSZ effect. 
In this letter we compute the imprint on CMB temperature anisotropies
due to the peculiar motions of gas in filaments and halos. 
The relative KSZ contribution with respect to the TSZ increases with decreasing 
temperature and the anisotropies generated could be used to trace
the gas distribution on large scales.  Briefly, in Sec. 2 we show 
the effect of individual filaments on the CMB temperature anisotropies.
We extend an early calculation of the KSZ probability distribution function due to halos 
by Yoshida, Sheth \&  Diaferio (2001) by including gas not bound in halos
and generalize the work of Juszkiewicz, Fisher \& Szapudi (1998). 
In Sec. 3 we compute the radiation power spectrum, showing that
it reaches its maximum at $l\sim 400$, an angular scale much larger than that of
clusters of galaxies; we also show that a statistical analysis 
including the anisotropies generated by the peculiar motions in the WHIM 
improves the fit to WMAP 3yr data.  Finally, we present our main conclusions.

\section{The effect of single filament.}

The temperature anisotropy due to the KSZ effect,
along a given line of sight ($los$), is the number density 
weighted average of the radial component of the peculiar velocity of electrons:
$\Delta T=T_o\sigma_T\int n_e(\hat{l})(v_{e,r}(\hat{l})/c) dl=\tau v_r/c$.
In this expression, $n_e$ is the electron number density,
$v_{e,r}$ is the projected peculiar velocity along the
line of sight $\hat{l}$, $\sigma_T$ is the Thomson cross section
and $c$ the speed of light. The last expression follows when
the KSZ signal is dominated by a single filament of size $L$ and radial component
of its peculiar velocity $v_r$. In this case, $\tau=\sigma_Tn_eL$ is the optical
depth along the filament of average electron density $n_e$.

To compute the probability distribution that a filament produces
a given KSZ anisotropy, we assumed that the matter distribution is
well described by a log-normal random field.
Coles \& Jones (1990) showed this approximation to be adequate when matter
density perturbations are weakly non-linear and the velocity
field is still in the linear regime. 
The probability that at any given point
and at redshift $z$ there is an overdensity $\xi=n_e(\vec{x},z)/\bar{n}_e(z)$, 
with $\bar{n}_e(z)$ the average electron density, is:
$f_1(\xi)=(\xi\sigma_\delta\sqrt{2\pi})^{-1}
\exp\left[-(\log\xi-\mu_\delta)^2/(2\sigma_\delta^2)\right]$
where $\mu_\delta$ and $\sigma_\delta$ are the mean and rms of the matter density contrast
in the linear regime (i.e., $\mu_\delta=0$). 
For filaments of characteristic scale length $L$, this distribution
is also the distribution of the optical depth $\tau$ at each location. 
The KSZ contribution includes the effect of
all baryons along the $los$, independently of whether they are in halos
or in the form of WHIM.

\begin{figure}[ht]
\plotone{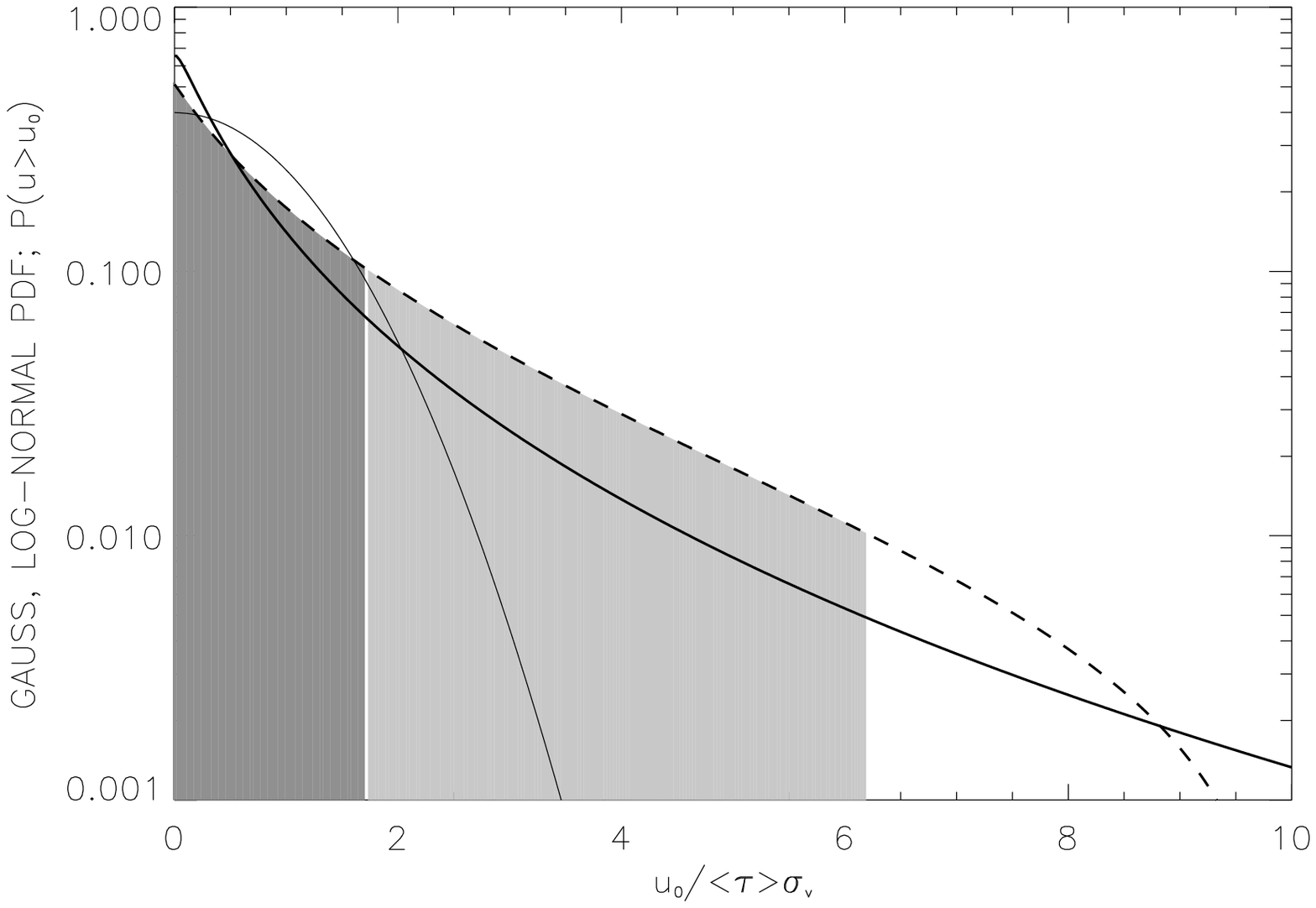}
\caption[]{KSZ (thick solid) and Gaussian (thin
solid) probability distribution functions. The dashed line represents 
$P(u>u_0)$, the probability that the a given filament produces a KSZ anisotropy
$u$ equal or larger than a given value $u_0$. Shaded areas correspond limit the regions of these
probabilities being 10\% and 1\%.
}
\label{fig:pdf}
\end{figure}

The velocity field
is accurately described as Gaussian Random Field with mean and rms $\mu_v,\sigma_v$.  
Its coherence scale is much larger than that of the density field;
the motion of galaxies, clusters and the IGM filaments is not
determined solely by the matter distribution in their vicinity,
but from a much larger scale. Under this assumption, we can assume
that density and velocity fields are independent. At each location, the 
probability distribution function (PDF) that an ionized gas cloud 
would produce a KSZ  of a given 
amplitude is simply the product of the distribution of the optical
depth (log-normal) and of the velocity (gaussian).
If $X=\tau$ and $Y=v_r$ are two continuous random variables with 
joint probability $f_{X,Y}(x,y)$, then the PDF of $U=XY=\tau v_r$ is:
$f_U(u)=\int f_{X,Y}(x,u/x)|x|^{-1}dx$ (Rohatgi, 1976).
When both random variables are independent, then $f_{X,Y}=f_Xf_Y$.
The resulting PDF is symmetric with respect to the y-axis.
In Fig.~\ref{fig:pdf} we plot only the region $u>0$; the thick solid line 
represents the PDF of the KSZ effect due to the IGM. 
For comparison, we also plot as a thin solid line the gaussian distribution. 
Both distributions were normalized such that the velocity and linear density 
fields have zero mean and unit variance ($\sigma_v=\sigma_\delta=1$). 
Then, u is expressed in units of $\langle\tau\rangle\sigma_v$, where $\langle\tau\rangle$
is the average optical depth of filaments of a size $L$. 
Let us remark that the KSZ distribution has much larger tails than the 
normal distribution.  The same PDF was computed by Yoshida et al. (2001) for the halo model.
In their approach, it was unlikely that a given $los$ will contain more than one halo and
their PDF was dominated by the more abundant and less massive halos,
and only in the tails of the distribution is the contribution of the more
massive halos significant. In ours, the optical depth along each line of sight can be
much larger since it includes the effect of baryons in halos together
with the WHIM in low dense regions, and this effect will raise the 
tails of the PDF. 

More informative is the Distribution Function $F(u)=\int_{-\infty}^{u} f_U(u')du'$.  
In Fig.~\ref{fig:pdf} we plot the quantity $P(u>u_0)=1-F(u_0)$ (dashed line) 
that represents the probability that a given filament produces a KSZ signal $u$ equal
or larger than $u_0=\langle\tau\rangle v_r$.
The shaded regions $[0,1.7]$ and $[0,6.2]$ correspond to percentiles 0.9 and 0.99.
For a gaussian distribution, the probability that a random variable has an
amplitude 2.7 times the rms deviation is  1\%.
For the KSZ distribution, at the same percentage, the amplitude is 6.2 times the average;
if a filament of size $L$ and electron density $n_e$ has 
an optical depth $\tau\simeq 2\times 10^{-3}(n_e/10^3m^{-3})(L/1Mpc)$ and
moves with a velocity of $300km/s$, it will typically give rise to 
$\Delta T_{KSZ}\sim 5.5 \mu$K. About 9\% of the filaments in this
population will have enough velocity and density to
produce an effect in the range $[9.3,34]\mu$K and 1\% will
have values larger than $34\mu$K. The same region will produce, on average, a
thermal SZ signal of $\delta_{TSZ}\sim 11.5 G(\nu)\mu$K$(T_x/1keV)$,
where $G(\nu)\sim -2$ in the Rayleigh-Jeans regime.  

Observations with the Very Small Array
of the Corona Borealis Supercluster (G\'enova-Santos et al. 2005) have detected
an extended ($\sim 30$ arcmin) cold spot without significant X-ray emission on the 
ROSAT-R6 map, termed decrement H. Those authors described
the likelihood that it could be partly generated by an unknown massive 
cluster in the background or by a filament. They did not include
the significant KSZ contribution in their study. For example, 1\% of all the filaments 
of size $\sim 5$Mpc, aligned with the $los$ and with a coherent flow at similar scale 
will produce a $\delta_{TSZ}+\delta_{KSZ}\sim (-150,-200)\mu$K,
enough to explain the central temperature of $-230\pm 23\mu$K of decrement H.
Further observations, like those to be carry out by the PLANCK satellite, at various frequencies
and with better angular resolution, are required to quantify the KSZ contribution.

\section{Contribution to CMB anisotropies.}

The KSZ effect is independent of redshift; along a $los$
contributions from individual filaments will average out since their
peculiar velocities will have random directions. 
Since the non-linear density and velocity fields
evolve with redshift, so does the contribution of the filament network.
To estimate the temperature anisotropies generated by the
peculiar motions of filaments and halos
we adapted the formalism developed by Choudhury, Padmanabhan \& Srianand (2001) 
and ABM.  The correlation function of the KSZ signal is
\begin{equation}
C(\theta)=\langle\frac{\Delta T_{KSZ}(\hat{x})}{T_o}\frac{\Delta T_{KSZ}(\hat{x}')}{T_o}\rangle=
\frac{\sigma_T^2}{c^2}\int_o^{l(z_f)}\int_o^{l(z_f)} dl dl' \langle n_e(l\hat{x},z)v_{e,r}(l\hat{x},z)
n_e(l\hat{x}',z')v_{e,r}(l\hat{x}',z')\rangle
\label{eq:definition}
\end{equation}
where $\theta$ is the angular separation between the two $los$, $\hat{x}$ and $\hat{x}'$.
The double integration is carried out from the origin to the highest redshift $z_f$ where the
IGM is still fully ionized. 

The main difficulty in making an analytical estimate 
is to relate density and velocity fields at each location
since both random fields have very different scale lengths.
If we assume they are independent, like in the previous section, we can 
decompose the velocity field into two contributions: the random motion
of matter with respect to the center of mass $v_R(z)$,
and the velocity of the center of mass with respect to
a comoving observer, also termed Bulk Flow $V_B$: $v_e(l\hat{x},z)=V_B(z)+v_R(l\hat{x},z)$. 
With this decomposition, the integrals containing the random component 
in eq.~(\ref{eq:definition}) average out and 
\begin{equation}
C(\theta)=\frac{\sigma_T^2}{c^2}
\int_o^{l(z_f)}\int_o^{l'(z_f)} dl dl' V_B(R(z))D_v(z)V_B(R(z'))D_v(z')
\langle n_e(l\hat{x},z)n_e(l\hat{x}',z')\rangle
\label{eq:corr}
\end{equation}
where $V_B(R(z))$ denotes the mean bulk velocity of a sphere with radius 
$R(z)$, the comoving distance from the observer to a filament at redshift $z$ and
$D_v(z)$ is the velocity linear growth factor.

Assuming a log-normal distribution for matter and baryons, 
in the non-linear regime, the electron number
density $n_e({\bf x},z)$ in the IGM is: 
$n_B({\bf x},z)=n_0(z)\exp[\delta_B({\bf x},z)-\Delta_B^2(z)/2]$,
where ${\bf x}$ denotes the spatial position at redshift $z$
and $|{\bf x}(z)|$ is the proper distance;
$\delta_B({\bf x},z)$ is the linear baryon density contrast, $n_0(z)=\rho_B(1+z)^3/\mu_B m_p$
and $\rho_B$, $m_p$ are the baryon density and proton mass, respectively. Finally,
\be
\Delta_B^2(z)={\langle {\delta_B^2({\bf x},z)}\rangle} = 
D^2(z)\int{\f{d^3k}{(2\pi)^3}\f{P_{DM}(k)}{[1+x_b^2(z)k^2]^2}} ,
\e
where $D(z)=D(z,\Omega_{\Lambda},\Omega_m)$ is the matter linear growth factor.
The spatial average in eq.~(\ref{eq:corr}) is ABM:
\begin{equation}
C(\theta)=\frac{\sigma_T^2}{c^2}
\int_o^{l(z_f)}\int_o^{l'(z_f)} dl dl' V_B(R(z))D_v(z)V_B(R(z'))D_v(z')
n_e(z)n_e(z')F(\theta,z,z'),
\label{eq:corr_simple}
\end{equation}
with
\be
F(\theta,z,z')=-1 + \exp\left[\frac{D(z)D(z')}{2\pi^2}
\int^\infty_0{{P(k)k^2 dk\over [1 + x_b^2(z)k^2][1 + x_b^2(z')k^2]}{\sin
    (k|{\bf x}-{\bf x'}|)\over k|{\bf x}-{\bf x'}|}}\right] ,
\e
where $|{\bf x}-{\bf x'}|$ denotes the proper distance between two filaments at
positions ${\bf x}(z)$ and ${\bf x'}(z')$ separated by the angle $\theta$,
$P(k)$ is the matter
power spectrum and $x_b$ is the Jeans length at each redshift, that relates
the matter power spectrum to that of baryons (Fang et al. 1993):
$x_b(z)=H_0^{-1}\left[(2 \gamma k_{\rm B} T_m(z))/(3 \mu m_p \Omega_m (1+z))\right]^{1/2}$.
In this last expression, $T_m$ is the mean
temperature of the IGM, $\gamma\sim(1.2-1.6)$ is the polytropic index, 
$\Omega_m$ is the cosmological fraction of matter density and
$\mu$ is the mean molecular weight of the IGM. The helium weight fraction
was $Y=0.24$.  The factor $\gamma T_m$ defines the Jeans length and fixes
the smallest scale that contributes to the spectrum. We will take 
$\gamma T_m \approx 2\times 10^4$K (Schaye et al. 2000).
The number density of electrons $n_e$ in the IGM can be obtained by assuming ionization
equilibrium between recombination, photo-ionization and collisional ionization. At the 
conditions valid for the IGM (temperature in the range $10^4-10^7$K,
and density contrast $\delta < 100$) the degree of ionization is very high. 
Commonly $n_e = \epsilon n_B$, with $0.9<\epsilon\le 1$ depending on the 
degree of ionization. For future reference, $x_b=0.3-0.5h^{-1}$Mpc, depending
on model parameters (see ABM).

In eq.~\ref{eq:corr_simple}, for each $\theta$ and $z$ the function 
$F(\theta,z,z')$ is strongly peaked at $z = z'$ and is different from zero
only within a very narrow range $z' \in [z-\Delta z, z+\Delta z]$. 
The effective peak width $2\Delta z$ increases and its maximum amplitude 
decreases with decreasing angle and increasing redshift, 
simplifying the numerical evaluation of eq.~\ref{eq:corr_simple};
in our numerical integration we took $\Delta z=0.1$. 
Integration around the peak width was carried out with high accuracy, using $z,z'$ intervals of 
$\delta z=5\times 10^{-4}$ and angle separations of 30 arcseconds.
Further integration (with respect to $z$ only) is carried out with a much smoother function 
and can be performed by a relatively rough spacing in redshift. 

\begin{figure}[ht]
\plotone{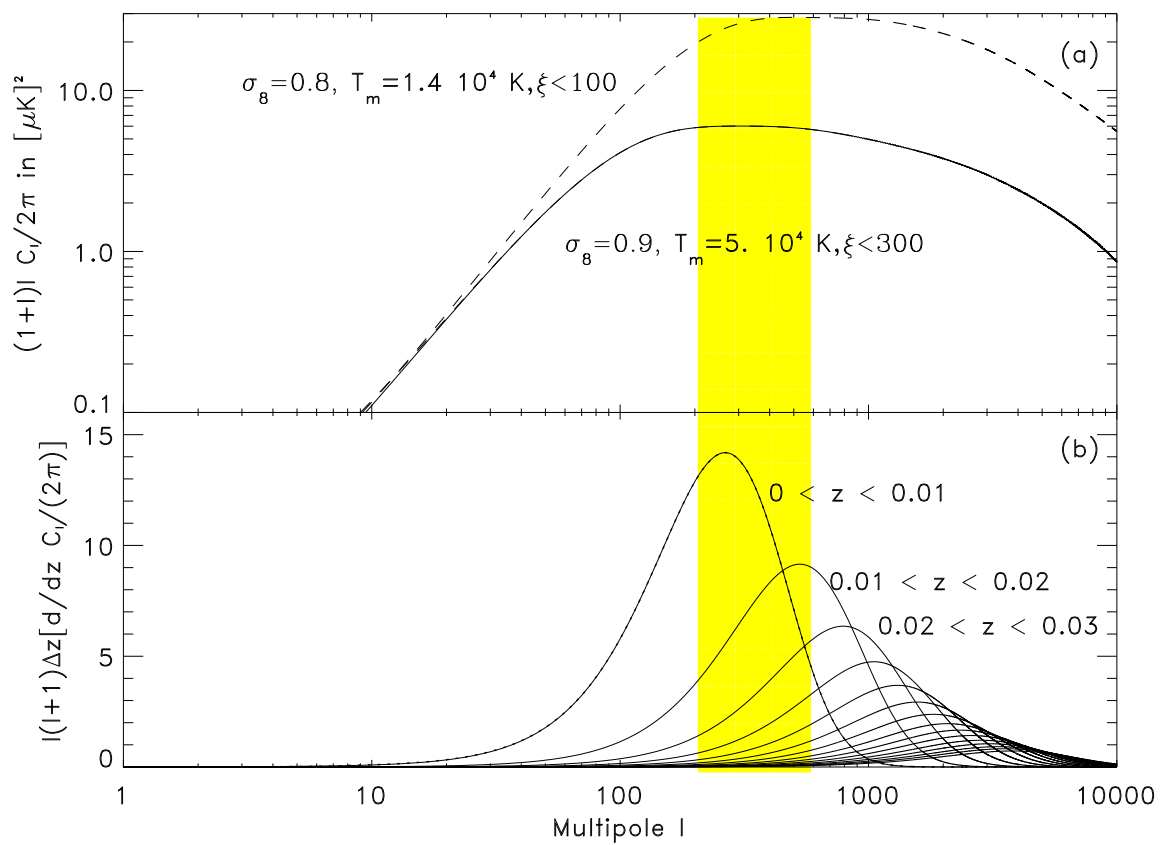}
\caption[]{
(a) KSZ Radiation power spectra for different parameter sets $\sigma_8, 
T_m, \xi$ (see labels). (b) Differential Power spectrum contribution from 
subsequent redshift bins of width $dz=0.01$. The shaded area indicates the 
interval where power spectra with different parameters reach
their maxima.}
\label{fig:cl}
\end{figure}

In Fig.~\ref{fig:cl}a, we show the radiation power spectrum obtained using
the formalism developed before. Since some of the parameters in our model are poorly 
determined, we considered the effect on the radiation power spectrum of: 
(a) the amplitude of the matter power spectrum measured by $\sigma_8$, 
(b) the averaged temperature of the IGM, $T_m$ and 
(c) the maximum overdensity $\xi$ for which the log-normal
distribution is a good description of the IGM. 
We took $\sigma_8=(0.8;0.9),
T_m = (1.4\times 10^4; 5.\times 10^4)$ K and $\gamma=1.3$ throughout. We
restricted all integrations to overdensities $\xi \le 500$.
The overall shape of the power spectra is very similar for 
a large range of the parameter space: 
the spectrum peaks around $l\sim 200-600$ (shaded area)
with a maximum amplitude of $10-25\mu$K.
This signal is much larger than the one considered by Hajian et al (2007),
associated to the free electrons in the halo of the Galaxy.
The amplitude and angular scale depend on $x_b$ or, equivalently, on 
$T_m$. The Jeans length fixes the scale where the power in the baryon density
field falls below that of the DM. Decreasing $x_b$ decreases the scale 
where baryons evolve non-linearly and increases the fraction of baryons
sharing the same coherent motion, increasing the KSZ power spectrum
and shifting the maximum to smaller angular scales.
The power also increases with  increasing either $\sigma_8$ or $\xi$.
If we vary parameters within the range $\sigma_8\in (0.7-0.9)$ and $\xi\in(50-500)$
the amplitude of the power spectrum could change up to a factor $\approx 5$.
$T_m$ has the largest effect: the amplitude scales as $\sim T_m^{-6}$ similarly 
to the Thermal Sunyaev-Zeldovich contribution (see ABM).

In Fig.~\ref{fig:cl}b we plot the contributions to the power spectrum from 
different redshift bins ($0< z < 0.15; dz = 0.01$).  As
remarked above, the differential contribution strongly decreases with redshift.
The signal is dominated by: (1) the bulk motion on scales
of $300h^{-1}$Mpc and below  and (2) by the structures
with largest density contrast.  Contributions from higher
redshifts and lower densities are negligible due to the superposition
of filaments with randomly oriented velocities with decreasing contributions.
Like in our earlier results for the TSZ effect,
the scale of the maximum is fixed by the Jeans length; at z=0.01, $x_b\sim 400h^{-1}$kpc
corresponds to an angular scale $l\sim 300$, as shown. 

Comparing the spectrum of Fig.~\ref{fig:cl} with the results of
the KSZ effect obtained using the halo model (Molnar \& Birkinshaw 2000,
Cooray \& Sheth 2002) the former shows more power (factor 3-10, depending on model
parameters) and at different angular scale. While the halo model
only accounts for the gas in halos, our formalism also incorporates the
gas in low dense regions. Those regions with intermediate density contrast
include the largest gas fraction at low redshift (see ABM)
when the bulk flow velocities are largest, which boost the power 
with respect to the halo model result.

Let us consider whether the KSZ contribution described above
could explain the excess of power in the range $l\sim 200-600$ measured 
by WMAP 3yr data with respect to the best fit model
(see Hinshaw et al 2007, Fig. 17). We explored the parameter
space of $\Lambda$CDM models using a Monte Carlo Markov Chain (MCMC);
we added a KSZ contribution with an amplitude at the maximum of $25\mu$K
to the radiation power spectrum of each cosmological model. 
This approach is not self consistent since when $\sigma_8$ changes so does
the amplitude of the KSZ contribution. Even though, the best fit model improved
the fit in one unit: $\Delta\chi^2=1$. This improvement is not statistically
significative from a Bayesian point of view (Liddle 2004): we introduced two parameters
$\gamma T_m$ and the maximum overdensity $\xi$, to obtain only a mild improvement on the fit.
Comparison with the concordance model, parameters deduced from
our MCMC differ by less than 1\%, being the variation on $\Omega_bh^2$ the most 
significative. This exercise does show that ionized gas could originate measurable CMB
temperature anisotropies and CMB observations could provide evidence of the
gas distribution at large scales.

\section{Discussion.}

If the 'missing baryons' are in the form of a highly ionized WHIM,
and this medium shares the same velocity field than galaxies and dark matter,
then it will produce a contribution to the CMB temperature anisotropies
that could be detectable.  Gas in halos and filaments with bulk flows aligned with
the $los$ could explain a significant fraction of the signal measured in the decrement H
of the Corona Borealis supercluster observed by the VSA interferometer.
Also, we computed the effect of the overall population of filaments on
CMB temperature anisotropies. Their contribution has a caracteristic
angular scale at $l\sim 400$, very different from that of clusters
of galaxies. Uncertainties in
our model parameters translate into an uncertainty of a factor $5$
in the amplitude of the power spectrum.

Intrinsic and KSZ temperature anisotropies have the same frequency dependence, they
can not be separated directly from the data.  Using a MCMC analysis
we have shown that this extra component improves the fit of the
concordance model to WMAP 3yr data, but the improvement can not
be considered significant. WMAP 5yr data or the future
PLANCK observations could have enough statistical power to indicate the existence
of this component. Further, PLANCK has enough angular resolution and frequency coverage
as to measure the KSZ contribution of structures such as decrement H in the VSA data of
the Corona Borealis Supercluster.

\acknowledgments
F.A.B. acknowledges financial support from the Spanish Ministerio de Educaci\'on y Ciencia
(project FIS2006-05319) and Junta de Castilla y Le\'on (project SA010C05).
We thank R. Sheth for discussions. J.P.M. thanks the University of Salamanca for
its support and hospitality.

\clearpage

\end{document}